# Small and Finite Inertia in Stochastic Systems: Moment and Cumulant Formalisms

Denis S. Goldobin[1,2,a)] and Lyudmila S. Klimenko[1,2,b)]

[1]*Institute of Continuous Media Mechanics, UB RAS, Perm, Russia*
[2]*Perm State University, Perm, Russia*

[a)]Denis.Goldobin@gmail.com
[b)]Corresponding author: lyudmilaklimenko@gmail.com

**Abstract.** We analyze two approaches to elimination of a fast variable (velocity) in stochastic systems: moment and cumulant formalisms. With these approaches, we obtain the corresponding Smoluchovski-type equations, which contain only the coordinate/phase variable. The adiabatic elimination of velocity in terms of cumulants and moments requires the first three elements. However, for the case of small inertia, the corrected Smoluchowski equation in terms of moments requires five elements, while in terms of cumulants the same first three elements are sufficient. Compared to the method based on the expansion of the velocity distribution in Hermite functions, the considered approaches have comparable efficiency, but do not require individual mathematical preparation for the case of active Brownian particles, where one has to construct a new basis of eigenfunctions instead of the Hermite ones.

## INTRODUCTION

Inertia, even small, significantly increases complexity of collective behavior in active media with local and global interactions and oscillator networks [1–4]. For stochastic systems the limit of vanishing inertia requires a subtle treatment, as one cannot omit the highest order time-derivative in the presence of a delta-correlated forcing [5–9]. The derivation of the leading order corrections owned by a weak inertia is even more challenging and demanded for many stochastic systems, especially for Brownian particles (passive [5, 9] and active [10–14]).

The inertia-related increase of complexity is especially pronounced for populations of phase elements, which, in the absence of inertia, obey the Watanabe–Strogatz [15–18] and Ott–Antonsen theories [19, 20]. Recently [21], a circular cumulant approach was introduced for dealing with the systems where the applicability conditions of the Ott–Antonsen theory are violated. Within the framework of the circular cumulant formalism [21–24], one can consider weak inertia as a perturbation to the Ott–Antonsen properties and construct a low-dimensional description of the macroscopic collective dynamics of populations of phase elements. This task however can be accomplished in many different ways and, therefore, preliminary analysis of the moment and cumulant expansions with respect to a fast variable (velocity) is desirable.

The approaches of moment and cumulant expansions can be also beneficial for studies on Brownian particles, especially the active ones. In the latter case, different auto-propulsion laws require the construction of unique bases of special functions for the velocity distribution instead of the Hermite function basis [1, 5], which is specific to the case of passive Brownian particles.

In this paper we will deal with a system with inertia, governed by the following Langevin equation:

$$\mu\ddot{\varphi} + \dot{\varphi} = F(\varphi, t) + \sigma\xi(t), \quad \mu \ll 1, \qquad (1)$$

where $\mu$ is mass or a measure of 'inertia' in the system (for superconducting Josephson junctions [25, 26], power grid models [27, 28], etc.), $F$ is a deterministic force, $\sigma$ is the noise strength, $\xi$ is a normalized $\delta$-correlated

Gaussian noise: $\langle \xi \rangle = 0$, $\langle \xi(t)\xi(t') \rangle = 2\delta(t-t')$.

The corresponding Fokker–Planck equation for the probability density $\rho(v,\varphi)$, where $v = \dot\varphi$, reads

$$\partial_t \rho = -v \partial_\varphi \rho + \partial_v \left[ \frac{1}{\mu}(v - F(\varphi,t))\rho \right] + \frac{\sigma^2}{\mu^2} \partial_v^2 \rho \qquad (2)$$

(here $\varphi$ can be in a rotating reference frame, where it is useful).

The aim of this paper is to construct and analyze universal schemes for rigorous elimination of the velocity (or, generally, any fast variable) in the case of small/vanishing inertia and consider the effective dynamics solely for $\varphi$. We analyze two approaches to accomplishing this task:

- Moment formalism: representation in terms of moments $w_n(\varphi) = \int_{-\infty}^{+\infty} v^n \rho(v,\varphi)\,dv$;

- Cumulant formalism: representation in terms of $K_n(\varphi)$ (or $\kappa_n = K_n/n!$), defined via the moment- and cumulant-generating functions [29]:

$$f(s,\varphi) \equiv \sum_{n=0}^{\infty} w_n(\varphi) \frac{s^n}{n!}, \qquad \ln f \equiv \phi(s,\varphi) \equiv \sum_{n=0}^{\infty} K_n(\varphi) \frac{s^n}{n!}.$$

## Mathematical Preliminaries: Scaling Laws for Moments for Vanishing Inertia

In the Langevin equation we assume $\varphi = \langle\varphi\rangle + \tilde\varphi$, where $\langle ... \rangle$ stands for averaging over noise signal realizations. Hence,

$$\langle \dot\varphi \rangle = \langle F(\varphi,t) \rangle$$

and, neglecting obviously non-leading terms (since $\dot{\tilde\varphi} \gg F(\varphi,t) - \langle F(\varphi,t)\rangle$), one finds

$$\ddot{\tilde\varphi} + \frac{1}{\mu}\dot{\tilde\varphi} \approx \frac{\sigma}{\mu}\xi(t).$$

From the latter,

$$\dot{\tilde\varphi}(t) = \frac{\sigma}{\mu} \int_0^{+\infty} d\tau\, \xi(t-\tau) e^{-\frac{\tau}{\mu}},$$

which means that $\dot{\tilde\varphi}$ is a Gaussian random number. Let us find its variance;

$$\langle [\dot{\tilde\varphi}(t)]^2 \rangle = \frac{\sigma^2}{\mu^2} \int_0^{+\infty} d\tau_1 \int_0^{+\infty} d\tau_2\, 2\delta(\tau_1-\tau_2) e^{-\frac{\tau_1+\tau_2}{\mu}} = \frac{\sigma^2}{\mu}.$$

Thus,

$$\dot{\tilde\varphi} = \frac{\sigma}{\sqrt{\mu}} R,$$

where $R$ is a normalized Gaussian random number. Further,

$$\left\langle [\langle\dot\varphi\rangle(t)+\dot{\tilde\varphi}(t)]^n \right\rangle \approx \begin{cases} \left\langle [\dot{\tilde\varphi}(t)]^n \right\rangle, & \text{for even } n; \\ \left\langle [\dot{\tilde\varphi}(t)]^n \right\rangle + n\langle\dot\varphi\rangle\left\langle [\dot{\tilde\varphi}(t)]^{n-1} \right\rangle, & \text{for odd } n \end{cases} \sim \begin{cases} \dfrac{\sigma^n}{\mu^{n/2}}, & \text{for even } n; \\ n\langle\dot\varphi\rangle\dfrac{\sigma^{n-1}}{\mu^{(n-1)/2}}, & \text{for odd } n. \end{cases}$$

# FOKKER–PLANCK EQUATION

## Moment Formalism

One can introduce the moments for $v$:

$$w_n(\varphi) = \int_{-\infty}^{+\infty} v^n \rho(v,\varphi)\,\mathrm{d}v.$$

For these moments the Fokker–Planck equation (2) yields

$$\partial_t w_0 + \partial_\varphi w_1 = 0, \tag{3}$$

$$w_1 + \mu \partial_t w_1 = F w_0 - \mu \partial_\varphi w_2, \tag{4}$$

$$w_n + \frac{\mu}{n}\partial_t w_n = F w_{n-1} - \frac{\mu}{n}\partial_\varphi w_{n+1} + (n-1)\frac{\sigma^2}{\mu} w_{n-2} \quad \text{for } n \ge 2. \tag{5}$$

For constructing a regular perturbation theory with respect to small parameter $\mu$, it is convenient to take the scaling laws for $\langle v^n \rangle$ with respect to $\mu$ into account explicitly by means of rescaling

$$w_n = \begin{cases} \dfrac{1}{\mu^{n/2}} W_n, & \text{for even } n; \\ \dfrac{1}{\mu^{(n-1)/2}} W_n, & \text{for odd } n. \end{cases}$$

Then Eqs.(3)–(5) can be recast in a form free from $1/\mu$-coefficients:

$$\partial_t W_0 + \partial_\varphi W_1 = 0, \tag{6}$$

$$W_1 + \mu \partial_t W_1 = F W_0 - \partial_\varphi W_2, \tag{7}$$

$$W_n + \frac{\mu}{n}\partial_t W_n = \mu F W_{n-1} - \frac{\mu}{n}\partial_\varphi W_{n+1} + (n-1)\sigma^2 W_{n-2} \quad \text{for } n = 2m, \tag{8}$$

$$W_n + \frac{\mu}{n}\partial_t W_n = F W_{n-1} - \frac{1}{n}\partial_\varphi W_{n+1} + (n-1)\sigma^2 W_{n-2} \quad \text{for } n = 2m+1. \tag{9}$$

Rearranging the terms, one finds

$$\partial_t W_0 + \partial_\varphi W_1 = 0, \tag{10}$$

$$W_1 = F W_0 - \partial_\varphi W_2 - \mu \partial_t W_1, \tag{11}$$

$$W_n = (n-1)\sigma^2 W_{n-2} + \mu\left[ F W_{n-1} - \frac{1}{n}\partial_\varphi W_{n+1} - \frac{1}{n}\partial_t W_n \right] \quad \text{for } n = 2m, \tag{12}$$

$$W_n = (n-1)\sigma^2 W_{n-2} + F W_{n-1} - \frac{1}{n}\partial_\varphi W_{n+1} - \frac{\mu}{n}\partial_t W_n \quad \text{for } n = 2m+1. \tag{13}$$

This equation system contains only $\mu^0$- and $\mu^1$-terms; thus, it is suitable for dealing with the limit $\mu \to 0$.

*Adiabatic Elimination of Velocity*

System (10)–(13) for $\mu = 0$ yields

$$\partial_t W_0 + \partial_\varphi W_1 = 0, \tag{14}$$
$$W_1 = F W_0 - \partial_\varphi W_2, \tag{15}$$
$$W_n = (n-1)\sigma^2 W_{n-2} \quad \text{for } n = 2m, \tag{16}$$
$$W_n = (n-1)\sigma^2 W_{n-2} + F W_{n-1} - \frac{1}{n}\partial_\varphi W_{n+1} \quad \text{for } n = 2m+1. \tag{17}$$

Equation (16) yields

$$W_{2m} = (2m-1)!! \sigma^{2m} W_0.$$

From Eq. (17),

$$W_{2m+1} = 2m\sigma^2 W_{2m-1} + (2m-1)!! \sigma^{2m}(F - \sigma^2 \partial_\varphi)W_0.$$

With $W_2 = \sigma^2 W_0$, Eqs. (14)–(15) yield

$$W_1 = (F - \sigma^2 \partial_\varphi)W_0,$$

and

$$\partial_t W_0 + \partial_\varphi (F W_0) - \sigma^2 \partial_\varphi^2 W_0 = 0. \tag{18}$$

Thus, we obtain a conventional Fokker–Plank equation for $W_0$, and all $W_{n\geq 1}$ are trivially determined by $W_0$. Noteworthy, for deriving Eq. (18), it is sufficient to use Eqs. (14)–(16).

*Corrected Smoluchowski Equation ($\mu^1$-Correction)*

Let us derive the $\mu^1$-correction to Eq. (18). Keeping $\mu^1$-corrections for $W_0$, one can find from the infinite equation system (10)–(13):

$$\partial_t W_0 + \partial_\varphi W_1 = 0, \tag{19}$$
$$W_1 = F W_0 - \partial_\varphi W_2 - \mu \partial_t W_1, \tag{20}$$
$$W_2 = \sigma^2 W_0 + \mu\left[F W_1 - \frac{1}{2}\partial_t W_2 - \frac{1}{2}\partial_\varphi W_3\right], \tag{21}$$
$$W_3 = 2\sigma^2 W_1 + F W_2 - \frac{1}{3}\partial_\varphi W_4 + \mathcal{O}(\mu), \tag{22}$$
$$W_4 = 3\sigma^2 W_2 + \mathcal{O}(\mu). \tag{23}$$

Starting with substitution of $W_4$ into the expression for $W_3$, one can find step-by-step in Eq. (22), Eq. (21), and Eq. (20):

$$W_3 = 2\sigma^2 W_1 + FW_2 - \sigma^2 \partial_\varphi W_2 + \mathcal{O}(\mu),$$

$$W_2 = \sigma^2 W_0 + \mu\left[-\frac{\sigma^2}{2}\partial_t W_0 + (F - \sigma^2 \partial_\varphi)^2 W_0 - \frac{\sigma^2}{2}\partial_\varphi(FW_0) + \frac{\sigma^4}{2}\partial_\varphi^2 W_0\right] + \mathcal{O}(\mu^2),$$

$$W_1 = FW_0 - \sigma^2 \partial_\varphi W_0 + \mu\left[-(\partial_t F + F\partial_\varphi F)W_0 + \sigma^2(\partial_\varphi F)\partial_\varphi W_0\right] + \mathcal{O}(\mu^2).$$

Finally, to the $\mu^1$-order,

$$\partial_t W_0 + \partial_\varphi\left[\left(F - \mu(\partial_t F + F\partial_\varphi F)\right)W_0\right] - \sigma^2 \partial_\varphi\left[(1 - \mu \partial_\varphi F)\partial_\varphi W_0\right] = 0. \tag{24}$$

Thus, we obtain the corrected Smoluchowski equation [5, 9]. The effective Langevin equation (Stratonovich form) of Fokker–Planck equation (24) is

$$\dot{\varphi} = F - \mu\left(\partial_t + F\partial_\varphi + \frac{\sigma^2}{2}\partial_\varphi^2\right)F + \sigma\sqrt{1 - \mu\partial_\varphi F}\,\xi(t).$$

*High-Order Approximations*

The conventional adiabatic elimination of a fast variable requires first three moments $w_0$, $w_1$, $w_2$; the first correction for small $\mu$ requires $w_3$ and $w_4$. Running equation system (3)–(5) for $w_0$, $w_1$, ..., $w_{2m+2}$ with formal closure $w_{2m+3} = 0$ yields the order of accuracy $\mu^m$.

## Cumulant Formalism

To derive the Fokker–Planck equations in terms of cumulants, let us start with the equation system for $w_n$:

$$nw_n + \mu \partial_t w_n = nFw_{n-1} - \mu \partial_\varphi w_{n+1} + n(n-1)\frac{\sigma^2}{\mu}w_{n-2},$$

which in terms of $f(s,\varphi) = \sum_{n=0}^{+\infty} w_n \frac{s^n}{n!}$ acquires the following form:

$$(s\partial_s + \mu\partial_t)f = \left(sF - \mu\partial_s\partial_\varphi + s^2\frac{\sigma^2}{\mu}\right)f.$$

Recasting the latter system in terms of $\phi = \ln f$, one can notice $\partial f = f\partial\phi$ and obtain

$$(s\partial_s + \mu\partial_t)\phi = sF + s^2\frac{\sigma^2}{\mu} - \mu\left[\partial_s\partial_\varphi\phi + (\partial_s\phi)(\partial_\varphi\phi)\right],$$

which, in turn, with $\phi = \sum_{n=0}^{+\infty} K_n \frac{s^n}{n!}$ yields the required equation system in terms of cumulants;

$$\mu \partial_t K_0 = -\mu \left[ \partial_\varphi K_1 + K_1 \partial_\varphi K_0 \right], \tag{25}$$

$$(n + \mu \partial_t) K_n = F \delta_{1n} + \frac{2\sigma^2}{\mu} \delta_{2n} - \mu \left[ \partial_\varphi K_{n+1} + \sum_{j=0}^{n} \frac{n!}{j!(n-j)!} K_{j+1} \partial_\varphi K_{n-j} \right] \quad \text{for } n \geq 1. \tag{26}$$

For the ease of comparison to the formalism of circular cumulants [21, 24] we introduce $\kappa_n = K_n / n!$ and recast the latter equation system as

$$\mu \partial_t \kappa_0 = -\mu \left[ \partial_\varphi \kappa_1 + \kappa_1 \partial_\varphi \kappa_0 \right], \tag{27}$$

$$(n + \mu \partial_t) \kappa_n = F \delta_{1n} + \frac{\sigma^2}{\mu} \delta_{2n} - \mu \left[ (n+1) \partial_\varphi \kappa_{n+1} + \sum_{j=1}^{n+1} j \kappa_j \partial_\varphi \kappa_{n+1-j} \right] \quad \text{for } n \geq 1. \tag{28}$$

We notice, with Eqs. (25)–(26), the conventional elimination of a fast variable requires the first three cumulants or moments $w_0$, $w_1$, $w_2$ with Eqs. (3)–(5); the first correction for small $\mu$ requires $w_3$ and $w_4$. Let us find what the accuracy of different truncations with the cumulant equation chain is:

$$\partial_t K_0 = -\partial_\varphi K_1 - K_1 \partial_\varphi K_0, \tag{29}$$

$$(1 + \mu \partial_t) K_1 = F - \mu \left[ \partial_\varphi K_2 + K_1 \partial_\varphi K_1 + K_2 \partial_\varphi K_0 \right], \tag{30}$$

$$(2 + \mu \partial_t) K_2 = \frac{2\sigma^2}{\mu} - \mu \left[ \partial_\varphi K_3 + K_1 \partial_\varphi K_2 + 2K_2 \partial_\varphi K_1 + K_3 \partial_\varphi K_0 \right], \tag{31}$$

$$(3 + \mu \partial_t) K_3 = -\mu \left[ \partial_\varphi K_4 + K_1 \partial_\varphi K_3 + 3K_2 \partial_\varphi K_2 + 3K_3 \partial_\varphi K_1 + K_4 \partial_\varphi K_0 \right], \tag{32}$$

$$(4 + \mu \partial_t) K_4 = -\mu \left[ \partial_\varphi K_5 + K_1 \partial_\varphi K_4 + 4K_2 \partial_\varphi K_3 + 6K_3 \partial_\varphi K_2 + 4K_4 \partial_\varphi K_1 + K_5 \partial_\varphi K_0 \right], \tag{33}$$

$$\ldots .$$

*Corrected Smoluchowski Equation*

Let us compare the solution of equation chain (29)–(33) up to $\mu^1$-terms (for $\mu \ll 1$) to equation system (19)–(23). First of all, the scaling laws of divergence for $K_n$ are different from those of $w_n$: $K_0 \sim K_1 \sim \mu^0$, $K_2 = \mu^{-1} const + \mathcal{O}(1)$, $K_{n \geq 3} \sim \mu^0$. Equations (29)–(33) can be rewritten as

$$\partial_t K_0 = -\partial_\varphi K_1 - K_1 \partial_\varphi K_0, \tag{34}$$

$$K_1 = F - (\mu K_2) \partial_\varphi K_0 - \mu \left[ \partial_t K_1 + \partial_\varphi K_2 + K_1 \partial_\varphi K_1 \right], \tag{35}$$

$$K_2 = \frac{\sigma^2}{\mu} - (\mu K_2) \partial_\varphi K_1 - \frac{\mu}{2} \left[ \partial_t \left( K_2 - \frac{\sigma^2}{\mu} \right) + \partial_\varphi K_3 + K_1 \partial_\varphi K_2 + K_3 \partial_\varphi K_0 \right], \tag{36}$$

$$K_3 = -(\mu K_2) \partial_\varphi K_2 - \frac{\mu}{3} \left[ \partial_t K_3 + \partial_\varphi K_4 + K_1 \partial_\varphi K_3 + 3K_3 \partial_\varphi K_1 + K_4 \partial_\varphi K_0 \right], \tag{37}$$

$$K_4 = -(\mu K_2) \partial_\varphi K_3 - \frac{\mu}{4} \left[ \partial_t K_4 + \partial_\varphi K_5 + K_1 \partial_\varphi K_4 + 6K_3 \partial_\varphi K_2 + 4K_4 \partial_\varphi K_1 + K_5 \partial_\varphi K_0 \right]. \tag{38}$$

One can see the advantages of the cumulant representation: while $w_n \sim \mu^{-\text{floor}[n/2]}$, one finds $K_2 \sim \mu^{-1}$, $K_{n \neq 2} \sim \mu^0$. Moreover, the $\mu^1$-correction requires $w_3$ and $w_4$ in the terms of moments, while $K_2$ is sufficient with cumulants. We notice, the adiabatic elimination of velocity also requires $K_2$, i.e., the $\mu^0$- and $\mu^1$-approximations require the same number of cumulants: $K_0$, $K_1$, and $K_2$. For the $\mu^1$-approximation, Eqs. (34)–(38) yield

$$\partial_t K_0 = -\partial_\varphi K_1 - K_1 \partial_\varphi K_0,$$

$$K_1 = F - (\mu K_2)\partial_\varphi K_0 - \mu\left[\partial_t K_1 + \partial_\varphi K_2 + K_1 \partial_\varphi K_1\right] + \mathcal{O}(\mu^2),$$

$$K_2 = \frac{\sigma^2}{\mu} - (\mu K_2)\partial_\varphi K_1 + \mathcal{O}(\mu),$$

$$K_n = -(\mu K_2)\partial_\varphi K_{n-1} + \mathcal{O}(\mu) \quad \text{for } n \geq 3.$$

Further,

$$\partial_t K_0 = -(\partial_\varphi + K'_0)\left[F - \sigma^2 K'_0 + \mu(\partial_t F + F'F + \sigma^2 F'K'_0)\right] + \mathcal{O}(\mu^2), \tag{39}$$

$$K_1 = F - \sigma^2 K'_0 - \mu\left[\partial_t F + F'F + \sigma^2 F'K'_0\right] + \mathcal{O}(\mu^2),$$

$$K_2 = \frac{\sigma^2}{\mu} - \sigma^2 F' + \sigma^4 K''_0 + \mathcal{O}(\mu),$$

$$K_n = (-\sigma^2 \partial_\varphi)^{n-1}(F - \sigma^2 K'_0) + \mathcal{O}(\mu), \quad \text{for } n \geq 3.$$

One can see that the self-contained equation for the evolution of $K_0$ is more lengthy than Eq. (14) for $w_0$. Moreover, Eq. (39) is equivalent to Eq. (24), if one substitutes $K_0 = \ln W_0$ and notices that $\partial K_0 = W_0^{-1}\partial W_0$, $(\partial_\varphi + K'_0)(...) = W_0^{-1}\partial_\varphi(W_0(...))$.

## CONCLUSION

We have analyzed two different approaches to the description of the inertia effect in stochastic systems: moment and cumulant formalisms. Comparison of two approaches shows that, although cumulant equations (29)–(33) for finite $\mu$ are significantly more lengthy than equations for moments $w_n$, the convergence properties of $K_n$ for $\mu \to 0$ are better, than that of moments. The adiabatic elimination of velocity in terms of cumulants and moments requires the first three elements. However, the $\mu^1$-correction to the Smoluchowski equation in terms of moments requires five terms (see [9] for the multiple-dimension case), while in terms of cumulants $K_n$ the same first three elements $K_0$, $K_1$, $K_2$ are sufficient. Generally, for the $\mu^m$-correction one needs the leading order accuracy for $K_{m+1}$, i.e., the first $m+2$ cumulants are required. Meanwhile, in terms of $w_n$ (or $W_n$), one needs the first $2m+3$ moments. Compared to the method based on the expansion of the velocity distribution in Hermite functions [5, 1], the considered approaches have a commensurate efficiency, but do not require individual mathematical preparations for the case of active Brownian particles (e.g., [10]).

## ACKNOWLEDGMENTS

The authors are grateful to Arkady Pikovsky and Lutz Schimansky-Geier for useful discussions and comments. The work was supported by the Russian Science Foundation (grant No. 19-42-04120).